\documentclass[doublecol]{epl2} 

\usepackage{hyperref}
\usepackage{graphicx}
\usepackage{amsfonts} 
\usepackage{thmbox} 
\usepackage{empheq} 
\usepackage{version}
\usepackage{times}
\usepackage{subfigure}
\usepackage{textcomp}
\usepackage{makeidx}
\usepackage{enumerate}
\usepackage{amsmath}
\usepackage{fancyhdr}
\usepackage[top=20mm, bottom=30mm, left=15mm, right=15mm]{geometry}
\usepackage{color}\newcommand{\red}[1]{{\color{red}{#1}\normalcolor}}
\newcommand{\blue}[1]{{\color{blue}{#1}\normalcolor}}

\newcommand{\beq}{\begin{equation}}     \newcommand{\eeq}{\end{equation}}
\newcommand{\beqa}{\begin{eqnarray}}    \newcommand{\eeqa}{\end{eqnarray}}
\newcommand{\bde}{\begin{description}}  \newcommand{\ede}{\end{description}}
\newcommand{\ben}{\begin{enumerate}}    \newcommand{\een}{\end{enumerate}}

\newcommand{\inv}[1]{{\frac{1}{#1}}}

\newcommand{\inRbracket}[1]{{\left({#1}\right)}}

\renewcommand{\revision}[2]{{\blue{\null} \null{#2}}}
\renewcommand{\red}[1]{\null{#1}}
                                                                                                          
\title{Allosteric propagation of curvature along filament}
\shorttitle{Allosteric propagation of curvature along filament} 

\author{Ken Sekimoto\inst{1,2}}
\shortauthor{K. Sekimoto}

\institute{                    
  \inst{1} {Laboratoire Gulliver, UMR CNRS 7083, ESPCI Paris, Universit\'e PSL\\
		10 rue Vauquelin, 75005, Paris, France.}
  \\
  \inst{2} {Laboratoire Mati\`ere et Syst\`emes Complexes, UMR CNRS 7057, Universit\'e Paris Cit\'e,\\ 10 Rue Alice Domon et Léonie Duquet, 75013, Paris, France }
}

\abstract{Can a filament transmit the curvatures across the constituting modules and control them at one of its end? Inspired by the observation of protofilament --- constituent biopolymer of microtubule --- this question is addressed by a constructive approach. In our model a simple allosteric element in each module couples with the neighboring modules at its interfaces, which gives rise to a single degree of freedom to control the global shape of the filament.
The model can be analyzed in analogy with  discrete-time dynamical systems having a bifurcation of transcritical type.
}

\begin{document}

\maketitle

\section{Introduction}
One dimensional assemblies of identical modules can exhibit various intrinsic structures such as circle, spiral or helix apart from thermal fluctuations. Those structures are unique when the constituting module is rigid and the connection at the interface is inflexible. 
The protofilament, a linear assembly of tubulin dimers as constituting modules, has shown 
a wide distribution of curvature from sample to sample when the protofilaments are isolated, either as subcritical nucleus of a microtubule (oligomer) \cite{Ayukawa2021} or as the terminal part of the plus end of a microtubule \cite{McIntosh2018}. 

From the point of view of the mechanics with constraints, the above observations \null{motivates} us to design the possible mechanism by which the linear association of identical modules realises the propagation of the curvature whose magnitude is tuneable.
To simplify the problem, we limit ourselves to the two-dimensional curves as the global shape of the filament, excluding twist or helix. Also we ignore  thermal fluctuations and flexibility except for  free joints that constitute the elements of each module.
The requirement is that the system can be \revision{}{extended} by appending the module in the way that the system has always a single ``global shape variable'' in a continuous domain.
For these purposes it is clear that each module should bear at least one internal degree of freedom and that such freedom should correlate allosterically the interface between the neighboring modules. In fact a study of structural biologists \cite{KnossowN2004} suggests the correlation between the conformational change of the module interface and the curvature\footnote{See Fig.2c of \cite{KnossowN2004}}. Below we construct a toy model to verify the possibility of the above idea and \revision{}{examin} the properties of the model.

The organisation of the paper is as follows: In \S{\bf Real toy model} 
 we present a macroscopic construction by real wooden pieces and some bolts and nuts for demonstration of feasibility of the idea. 
Then in \S{\bf Numerical model} we introduce a mathematical model for the module (\S\S{\it Module and Interfaces}
), and define how the modules are connected (\S\S{\it Linkage and Dynamical System}
). Next, we present the results of numerical studies for some selected cases
\S\S{\it Case studies} 
 and then develop a more general view in terms of the discrete-time dynamical system in \S{\bf Normal form}. 
  The final section \S{\it Discussion} 
 is a summary and discussion.


\begin{figure}[h]
\centering
\subfigure[]
{\label{fig:elements}  \includegraphics[width=3cm]{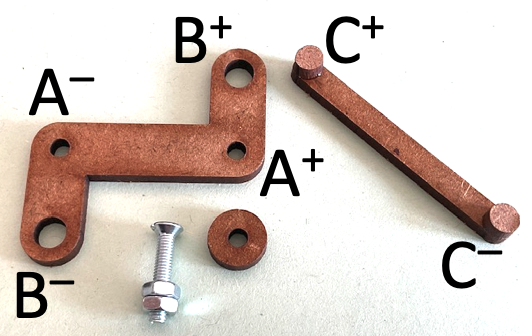}}
\hspace{2mm}
\subfigure[]
{\label{fig:perspective}    \includegraphics[width=5cm,angle=0.]{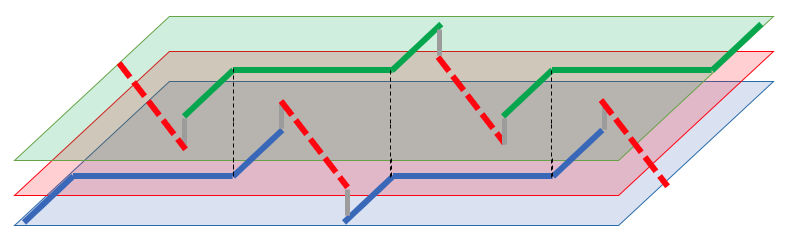}} 
\hspace{2mm}
\subfigure[]
{\label{fig:straight}    \includegraphics[width=8cm,angle=0.]{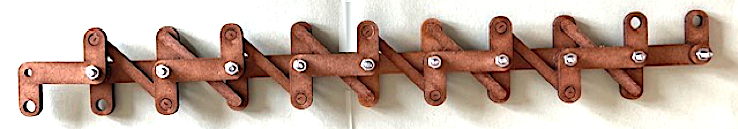}} 
\hspace{2mm}
\subfigure[]
{\label{fig:convex-real}    \includegraphics[width=4cm,angle=0.]{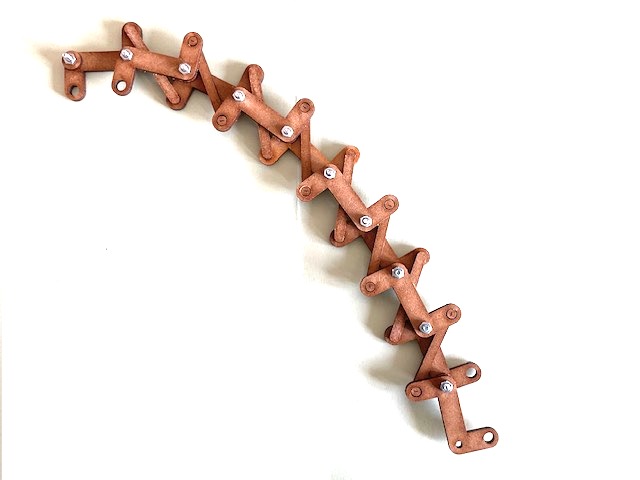}} 
\hspace{2mm}
\subfigure[]
{\label{fig:concave-real}    \includegraphics[width=4cm,angle=0.]{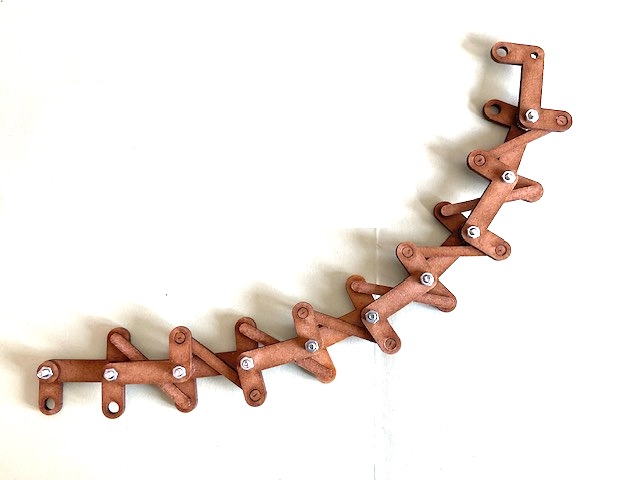}} 
\hspace{2mm}
\caption{(a) Constitutive elements of the module used in the toy model.  The Z-shaped ``backbone'' $B^-A^-A^+B^+$ and straight ``shaft'' $C^+C^-$ define the module, which are connected as described below. The ring, bolt and nut assure the free articulation at $A^\mp.$ The distance between $A^-$ and $A^+$ is  30mm. (b) Perspective view of the 3D design of the toy model. The structure is drawn in the reference (straight) state. The backbone elements (thick lines) are alternatively laid in the top and bottom layers, while only the shafts (dashed lines) are found in middle layer. The vertical lines traversing multiple layers indicate the free joints. (c) The model realized in the reference straight conformation. Convex (d) and concave (e) conformations.}
\label{fig:design} 
\end{figure}
\section{Real toy model} \label{sec:toymodel}
The real physical modelling helps our intuition on the one hand but also serves as a feasibility check of the local three-dimensional arrangement of real objects.
An implicit constraint is that the module should not be too sophisticated nor powered or controlled by some external source.

Our starting idea is to couple the two interfaces of a module through an allosteric mechanism. We know that the shearing of a square box $\Box{}$ inclines its vertical sides on its left and right, if the edges are connected by free joints. In order to make incline the vertical sides in the opposite orientations, we may revise the square box to make a twisted box ${\bowtie}.$ 
To make propagate such an anti-correlated inclination, we conceived a toy whose module consists of the elements shown in Fig.\ref{fig:elements}.
We call the pieces $B^-A^-A^+B^+$ and $C^-C^+$ the backbone and shaft, respectively. 
The three-dimensional architecture of the pieces is schematised in a perspective view, Fig.\ref{fig:perspective}, where we took as the reference state the straight conformation.  
If we assign the index $i$ for each module from the root ($i=0$)
towards the plus end ($i=m$) of the chain, 
the joints in Fig.\ref{fig:perspective} can be represented as
\beqa \label{eq:joints}
A^+_i &=& A^-_{i+1},\qquad (i=0,\ldots, m-1),
\cr 
B^+_i&=& C^+_{i+ 1},\quad
B^-_i=C^-_{i- 1},\qquad (i=1,\ldots, m-1),
\eeqa
where, for example, $A^+_i$ is the element $A^+$ of the $i$-th module. In the context of the growing phase of biofilament, the index $i$ would also mean the temporal order of modular attachment. Fig.\ref{fig:straight} shows the realisation of these connections in the reference state with $m=9.$\footnote{We think that the alternation of the layers of backbone is avoidable by realising the joint $A^+_{i-1}A^-_{i}$ in a single plane.}

In handling this chain-like object, we can verify that the system has (ideally) only a single continuous degree of freedom. 
In short the present architecture, therefore, can make the curvature propagate. 
Figs.\ref{fig:convex-real} and \ref{fig:concave-real} show, respectively, examples of convex and concave conformations. We will discuss the details in \S{\bf Numerical model} 
 in more general context.

\section{Numerical model} \label{sec:numodel}
\subsection{Module and Interfaces} 
\begin{figure}[h!!]
\subfigure[]
{\label{fig:5a} \includegraphics[width=4.cm,angle=0.]{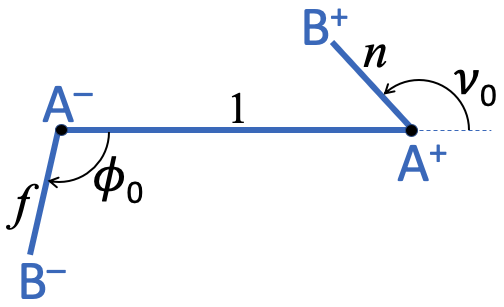}}
\hspace{-0mm}
\subfigure[]
{\label{fig:5x} \includegraphics
[width=2.5cm,angle=0.]{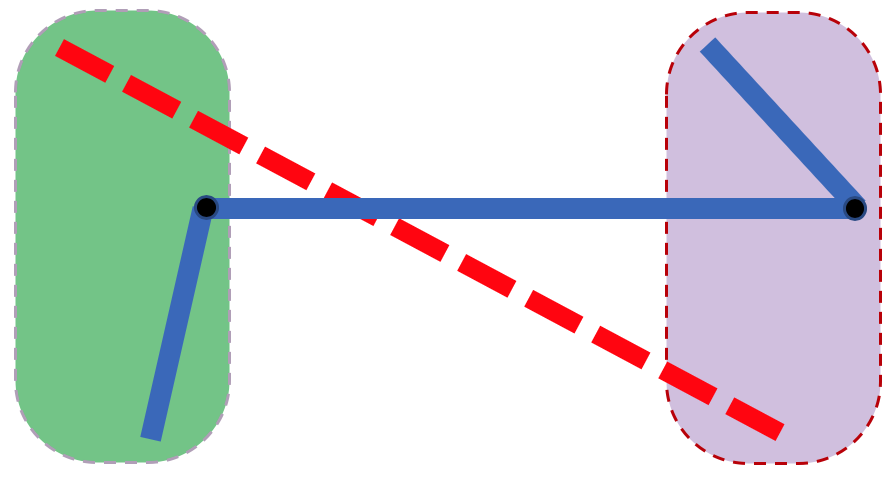}} 
\hspace{12mm}
\subfigure[]
{\label{fig:5y} \includegraphics[width=4.5cm,angle=0.]{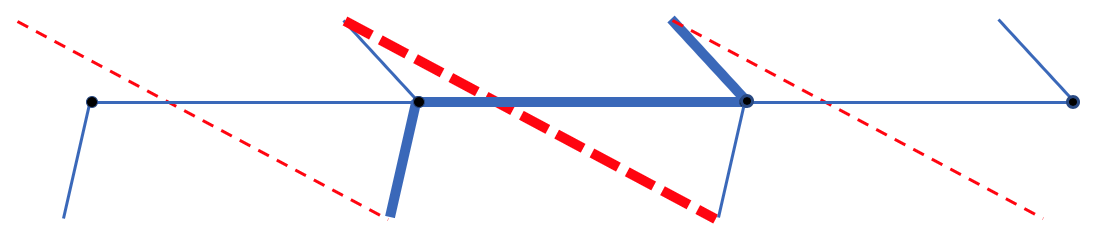}}
\hspace{5mm}
\subfigure[]
{\label{fig:5b} \includegraphics[width=3.2cm,angle=0.]{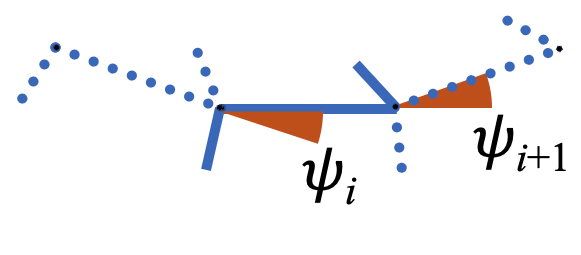}}\hspace{2mm}
\caption{(a) Backbone of each module and its parameterisation relative to the ``vertebrate,'' $\null{A^-\! A^+},$ of unit length. (The backbone in Fig.\ref{fig:elements} corresponds to the case of $f=n=1/2$ and $\nu_0=\phi_0=\pi/2.$)
(b) Backbone (solid lines) and shaft (dashed line) define two interfaces (shaded zones).
 (c) Determination of the length of shaft, $L_0,$ (the thick dashed line) through the 
 matching condition of the interfaces in the reference conformation.
(d) Assignment of the flexion angles 
\revision{\bf(Q1)\!\!}{We say $\psi_i>0$ 
when the $(i+1)$-th module is rotated counter-clockwise relative to the $i$-th module. When $\psi_i>0$ ($\psi_i<0$) for all $i$ we say that the global shape is concave (convex), respectively.} These angles are tightly connected through (\ref{eq:map-cond}) in the main text. 
 }\label{fig:geometrical-parameters}
\end{figure}

We characterise the shape of the individual backbone by the four real parameters, $\{f,\phi_0,n,\nu_0\}$ \revision{(Q2)\!\!}{with $0<\phi_0<\pi$ and $0<\nu_0<\pi,$, which we call ``module parameters,'' see} Fig.\ref{fig:5a}, 
where the central edge $A^-A^+$ is supposed to be of unit length.\revision{}{\footnote{
\revision{(Q2)\!\!}{The module parameters constitutes a set, which is a four-dimensional manifold. Later on we discuss the notion of ``global shape variable,'' which distinguishes different conformations of the linear chain. In the latter case the module parameters of the individual backbone modules are fixed.}}}
The interfaces with the neighbour modules are formed, as shown by the shaded regions in Fig.\ref{fig:5x}. The length of the shaft \revision{}{($L_0$)} is {\it not} the independent parameter but determined through the definition of the reference configuration, see below.

\subsection{Linkage and Dynamical System} 
The allosteric propagation of curvature through the filament is realised through the matching condition at the interfaces of modules. We impose that the aforementioned conditions (\ref{eq:joints}) are satisfied by the straight conformation.
The length of the shaft $L_0$ then obeys,
\beq \label{eq:L0}
L_0=
\left\|  \left(\!\! \begin{array}{c} 1\\ 0 \end{array}\!\! \right) \!+\!  f \! \left(\!\! \begin{array}{c} \cos(-\phi _0  {})\\ \sin(-\phi _0 )  \end{array}\!\! \right) \!-\!  n \! \left(\!\! \begin{array}{c} \cos(\nu _0  {} )\\ \sin(\nu _0   {} )   \end{array}\!\! \right)  \right\|,
\eeq
where $\left\|\cdot \right\|$ is the module of the vector , see Fig.\ref{fig:5y}.

Once the dimensions of elements are defined, we count the number of \revision{}{degrees of} freedom of the toy model, for example, shown in Fig.\ref{fig:straight}. There we see the eight shafts that constrain the nine flexion angles along the ten backbone elements. We, therefore, have a single (9-8=1) continuous degree of freedom.
To see how the flexion angles are constrained iteratively, we denote by ${\psi}_i$ the flexion angle at the $i$-th joint, 
see Fig.\ref{fig:5b}\revision{}{\footnote{\label{fn:A} 
\revision{(Q1)\!\!}{The assignment of the sign of $\psi_i$ is  linked to the assignment of the `time' direction, $i\to (i+1),$ see Appendix A\ref{sec:app-sign} } }}. 
Then the second line of (\ref{eq:joints}) under the fixed length $L_0$ of the shaft imposes the relation including ${\psi}_i$ and ${\psi}_{i+1}$ for $i=0,1,\ldots, m-1$:
{\small
\beqa \label{eq:map-cond}
&& \,\,\,\left\|  \left(\!\! \begin{array}{c} 1\\ 0 \end{array}\!\! \right) \!+\!  f \! \left(\!\! \begin{array}{c} \cos(-\phi _0 \!+\!  {{\psi}_{i+1}})\\ \sin(-\phi _0 \!+\!  {{\psi}_{i+1}})  \end{array}\!\! \right) \!-\!  n \! \left(\!\! \begin{array}{c} \cos(\nu _0 \!-\!  {{\psi}_i} )\\ \sin(\nu _0 \!-\!  {{\psi}_i} )   \end{array}\!\! \right)  \right\| 
\!=\!  L_{0} .
\cr && 
\eeqa}
The last equation implicitly defines a dynamical system,
\beq \label{eq:return-map} 
{\psi}_{i+1}=\Phi({\psi}_i),
\eeq 
with the fictive discrete time, $i.$

\subsection{Case studies} 
\begin{figure}[t ] %
\centering 
\subfigure[]
{\label{fig:figPIsur10e12e14}    \includegraphics[width=4.cm,angle=0.]{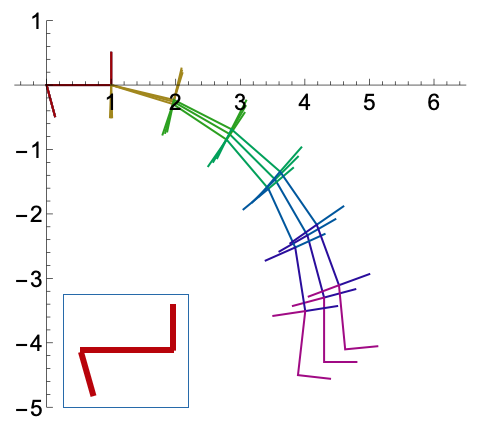}} 
\hspace{2mm}
\subfigure[]
{\label{fig:figPIsur11e12e13}    \includegraphics[width=4.cm,angle=0.]{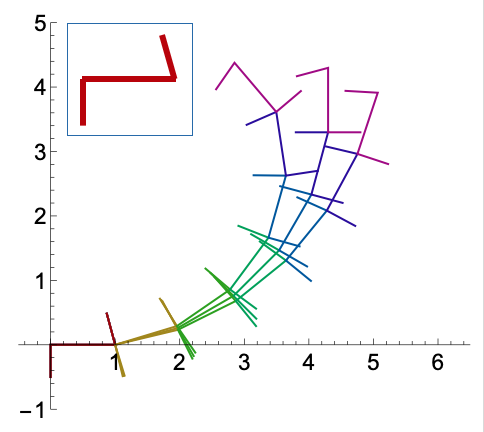}} 
\caption{Numerically calculated profiles with the flexion angles $\psi_i$ being in the vicinity of the fixed points of Eq.(\ref{eq:return-map}). The shafts are not shown.
Among the three profiles shown in (a) and (b) the middle ones are of the fixed-point evolution. The insets show the shape of backbone module studied with $f=n=1/2$ and (a)  $\{\phi_0,\nu_0\}=\{5\pi/12,\pi/2\}$ or (b) $\{\phi_0,\nu_0\}=\{\pi/2,7\pi/12\}.$ 
}
\label{fig:convergence-FP}
\end{figure}

\mbox{}\revision{}{Although} the modules are identical along the linear chain, the sequence $\{\psi_i\}$ generated by (\ref{eq:map-cond}) is in general non-constant. \revision{}{This is purely by a geometrical mechanism.}
\revision{(Q2)\!\!}{When the initial angle $\psi_0$ (at the leftmost joint $A_0^+=A_1^-$) is varied, 
the subsequent the deviations of $\psi_i$ ($i>0$) either decrease (Fig.\ref{fig:figPIsur10e12e14}) or increase (Fig.\ref{fig:figPIsur11e12e13}) with $i.$ 
That is, how the sequence $\{\psi_i\}$ evolves in `time' $i$ depends on the module parameters.
In Fig.\ref{fig:convergence-FP}, we have set $f=n=1/2$ in common, while 
$\{\phi_0,\nu_0\}=\{5\pi/12,\pi/2\}$ in Fig.\ref{fig:figPIsur10e12e14} and 
$\{\phi_0,\nu_0\}=\{\pi/2,7\pi/12\}$ in Fig.\ref{fig:figPIsur11e12e13}.
Once a set of the module parameters is fixed, the evolution is deterministic in obeying (\ref{eq:map-cond}). Then the global shape of the system can be specified by an arbitrarily chosen representative flexion angle, for example, by $\psi_0.$ See Appendix B\ref{sec:app-shape} for a little more details.  }

\begin{figure}[h!!] %
\centering
\subfigure[]
{\label{fig:fig8-S}    \includegraphics[width=2.cm,angle=0.]{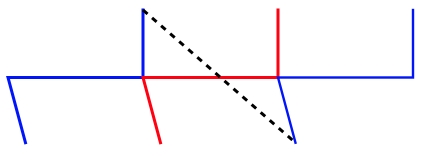}} 
\hspace{7mm}
\subfigure[]
{\label{fig:fig8-critical-non90}    \includegraphics[width=2.cm,angle=0]{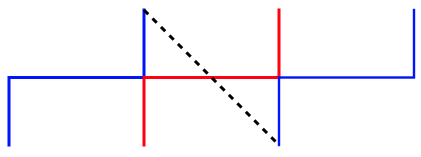}} 
\hspace{7mm}
\subfigure[]
{\label{fig:fig8-US}    \includegraphics[width=2.cm,angle=0.]{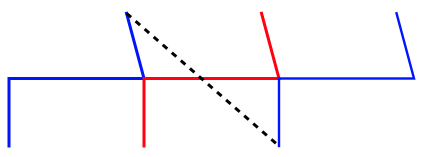}} 
\hspace{-4mm}
\subfigure[]
{\label{fig:fig6-S}    \includegraphics[width=2.7cm,angle=0.]{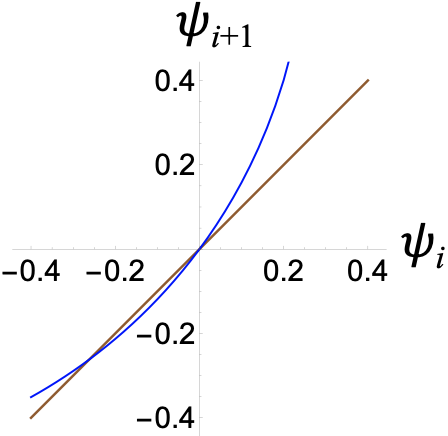}} 
\vspace{1mm}
\subfigure[]
{\label{fig:fig6-C}    \includegraphics[width=2.7cm,angle=0.]{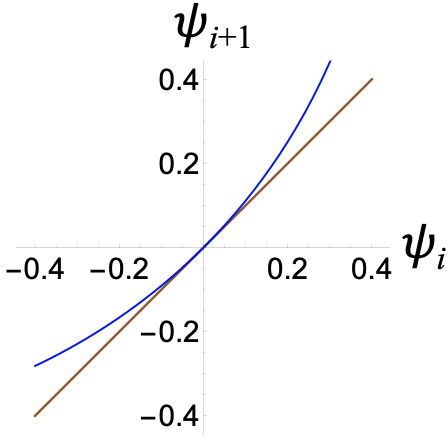}} 
\hspace{-0mm}
\subfigure[]
{\label{fig:fig6-US}    \includegraphics[width=2.7cm,angle=0.]{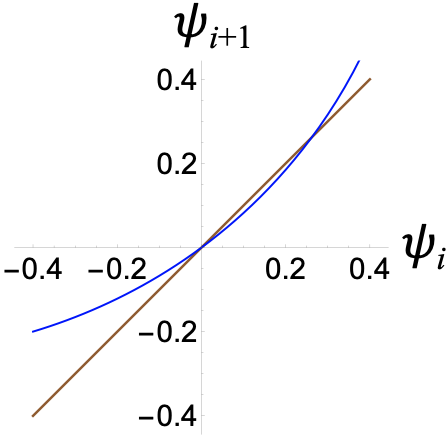}} 
\hspace{-3mm}
\subfigure[]
{\label{fig:transcritical} \includegraphics[width = 3.5cm]{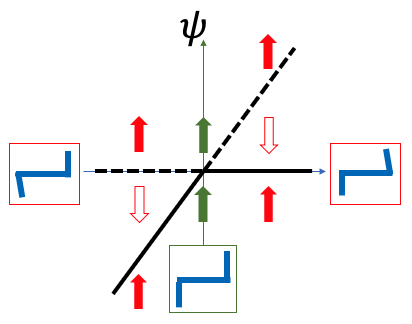}}
\caption{
Modular structure of backbone and shaft (dashed lines) in the reference state with
(a)  $(\phi_0,\nu_0)=(5\pi/12,\pi/2)$ realising a non-zero stable fixed point, $\psi=-{\pi}/{12},$ 
(b)  $(\phi_0,\nu_0)=(\pi/2,\pi/2)$ realising the marginal critical point, $\psi=0,$ and
(c)  $(\phi_0,\nu_0)=(\pi/2,7\pi/12)$ realising the unstable fixed point, $\psi={\pi}/{12},$  where $f=n=1/2$ in all cases.
(d)-(f): The return maps, $\psi_i\mapsto \psi_{i+1},$ 
\revision{}{calculated numerically through (\ref{eq:map-cond})}
with the backbones shown above (a)-(c), respectively. 
(g) Schematic diagram of the transcritical bifurcation whose normal form is given by Eq. (\ref{eq:normal}).
The horizontal axis represents the bifurcation parameter, $\mu,$ while the vertical direction represents the flextion angle $\psi$ with the arrows indicating the (discrete) flow from $\psi_i$ to $\psi_{i+1}.$ The solid [dashed] lines represent, respectively, the stable [unstable] fixed points. The origin ($\mu=\psi_i=0$) is the marginally stable\revision{}{, i.e., critical} fixed point.
The boxes indicate the correspondence between the return maps in Fig.\ref{fig:return-maps}\revision{}{(d-f)}  and the sign of $\mu.$ 
\revision{(Q5)\!\!}{Those red arrows, either filled or open, indicate the flow of $\psi_i$ with off-critical backbone shape, while the green arrows indicates the flow with critical backbone shape.} }
\label{fig:return-maps}
\end{figure}

\revision{(Q3+R2)\!\!}{
Regarding $\{\psi_i\}$ as a dynamical system evolving with discrete fictive time $i$, 
the numerical solution of (\ref{eq:map-cond}) allows to establish the return map (\ref{eq:return-map}) under a given set of module parameters. 
When we change the module parameters, the topological characteristics of the  return map changes, as shown in Figs.\ref{fig:return-maps}(a-f). These results indicates that there underlies a {\it transcritical} bifurcation. In very brief this category of bifurcation has always two fixed points of evolution, of which one is stable, i.e. convergent in its vicinity and the other is unstable, i.e., divergent in its vicinity. Across the bifurcation point these two fixed points pass each other, exchanging the stability/instability characters. For more details see for example \cite{Guckenheimer-Holmes,strogatz}. }

What we have studied in \S{\bf Real toy model}  belongs to the critical case, as depicted in Figs.\ref{fig:fig8-critical-non90} and \ref{fig:fig6-C}. 
In fact, Figs.\ref{fig:convex-real} and \ref{fig:concave-real} show the cases starting from  $\psi_{0}<0$ and of $\psi_{0}>0,$ respectively,\null{and the evolution that follows showed the monotonous decrement and increment of absolute curvature, in accordance with the critical return map Fig.\ref{fig:fig6-C}.\footnote{In the real toy model the conformations contain errors since the physical joints are not completely tight.}}


\section{Normal form}\label{sec:normalform} 
\mbox{}\revision{(Q3+R2)\!\!}{The normal form of bifurcation is a simplified return map that captures
the topological characteristics of the original return maps and the change among \red{them}.
While Figs.\ref{fig:return-maps}(d-f) are the numerical results based on (\ref{eq:map-cond}), their topological features are well reproduced by
the following standard analytical form, which is called the {\it normal form} of transcritical bifurcation:
\beq\label{eq:normal}
{\psi} _{i+1}={\psi}_i -\lambda {\psi}_i (\mu -{\psi}_i).
\eeq
Here $\lambda$ and $\mu$ are constant.In particular $\mu$ is the {\it bifurcation parameter} so that $\mu=0$ signifies the bifurcation/critical point. A sort of phase diagram corresponding to (\ref{eq:normal}) is given by Fig.\ref{fig:transcritical}. By definition this figure 
also captures the qualitative features of Figs.\ref{fig:return-maps}(d-f). 

Two mutually related points are to be discussed:
We present in the next paragraph a formal protocol to find all sets of module parameters for which the return map is critical, that is $\mu=0.$ Complementarily in Appendix C we present a derivation of the above normal form (\ref{eq:normal}) from (\ref{eq:map-cond}) under the hypothesis that $\{\psi_i,\psi_{i+1}, \mu\}$ are very small. There, the analysis also allows us to identify the critical manifold of $\mu=0$ in the shape space. Those readers not familiar with the transcritical bifurcation and its normal form are also invited to consult Appendix C\ref{sec:app-NF}. }

The presence of the critical point $\mu=0$ is a robust nature of this bifurcation, where the stable and unstable branches of fixed point crosses transversally. Having a single bifurcation parameter, $\mu,$ implies that the four-dimensional \revision{}{module parameter} space $\{f,\phi_0,n,\nu_0\}$ is divided into two domains separated by a three-dimensional manifold corresponding to the critical point $\mu=0,$ which we will denote by the equation, $\Lambda(f,\phi_0,n,\nu_0)=0.$

We can formally construct this manifold by noticing the {\it identity},
 $\Lambda(f,\phi_0,f,\pi-\phi_0)=0,$ telling us that any backbone with point-inversion symmetry realises the critical bifurcation system ($\mu=0$), see Fig.\ref{fig:fig6-C}, where $(\psi_i,\psi_{i+1})=(0,0)$ is the doubly degenerated fixed point. The above identity stems from the following equivalence relationship in (\ref{eq:map-cond}), 
\revision{(Q4)\!\!}{}
\beqa
 \{{\psi}_i\}_{0\le i\le m-1} && \mbox{\!\!\!\!\! \!\!\!\!\! \!\!\!\!\! 
 with $\{f,\phi_0,n,\nu_0\}$}
\cr 
\!\!\Leftrightarrow \quad \{ -{\psi}_{\revision{}{m}-i}\}_{0\le i\le m-1} &&  \mbox{\!\!\!\!\! \!\!\!\!\! \!\!\!\!\! 
with $\{n,\pi-\nu_0,f,\pi-\phi_0\}$},
\eeqa
where the stable fixed point is mapped into the unstable one and vice versa\footnote{{Since the filament generally has polarity, we distinguish a conformation and its mirror or point-inversion images in the 2D plane of the previous figures.}}.  Then 
$\Lambda(f,\phi_0,f,\pi-\phi_0)=0$ defines a two-dimensional sub-manifold of the entire critical manifold in the shape \revision{}{parameter} space. The former sub-manifold can be lifted to the latter manifold by solving $\Lambda(f,\phi_0,n,\nu_0)=0$ as a relation between $n$ and $\nu_0$ 
passing through $(n,\nu_0)=(f,\pi-\phi_0).$ 

\section{Discussion}
\label{sec:discussion}
We have presented a possible designing framework of statically propagating the curvature along an extendable chain of identical modules. The geometrical constraints imposed by the internal element (``shaft") of each module leaves only a ``global shape variable'',  the unique continuous degree of freedom that controls the global conformation of the filament. We note that this freedom,  a kind of zero-mode, does not need any {fine tuning}: 
\revision{(Q6)\!\!}{Even if the module parameters of individual backbone module are subject under static noises like $(f+\delta f_i,\phi_0+\delta \phi_i,n+\delta n_i,\nu_0+\delta \nu_i),$ we will still have a relation like (\ref{eq:map-cond}), leading to a `time-dependent' mapping function $\Phi(\psi_i,i)$ instead of $\Phi(\psi_i)$ in (\ref{eq:return-map}).\footnote{
\revision{}{On the other hand, the fixed points and their stability require a reconsideration in the presence of static noises.}} 
}
In real systems such a zero-mode will have finite persistence length due to the elastic deformability and thermal noise acting on the constituting elements of the module.
The architecture shown in Fig.\ref{fig:design} may remind us of the ``Ultra-Hand''\footnote{The ``Ultra-Hand'' is a toy invented by Gunpei Yokoi of Nintendo (1966, Japan), see Wikipedia: \url{https://en.wikipedia.org/wiki/Ultra_Hand}}, an adjustable reach extender whose global shape variable controls the affine extension of panthagraphs connected in series.

It is evidently far-fetched to expect that the present toy model gives some direct relevance to the states and growth of real biopolymers. We should be content, however, if  some of the aspects we have observed in the study of this model were to reveal certain universal features.
This study  brought us at least two conclusions: First, the geometrical construction imposes in general a non-uniform flexion/curvature except for some special angle of flexion. Secondly, if the modules are identical, the return map (\ref{eq:return-map}) can have \revision{}{a} non-trivial fixed point corresponding to a circular arc-shaped with specific curvature.

We have described the shaft of each module (Fig.\ref{fig:design}) as an allosteric agent that transmit the information of an interface to the other through the body of the module. From the viewpoint of the constrained mechanics, however, this
can be regarded as a non-local mechanical coupling between the backbones of next  nearest neighbours. Such structure reminds us of the epaxial muscles of a snake that are reported to have interlinked attachment sites \cite{snake-muscle}. While the shafts having fixed length serve to propagate the curvature through the zero-mode, the muscles of snake can vary the local curvature in space and time through their contraction.

\acknowledgements
The author thanks {\it Fablab} of Universit\'e Paris Cit\'e for letting him to use their laser cutter  to make the real toy model, and Olivier Marande for the technical suggestions and assistance. He thanks Etsuko Muto for critical comments. He also thanks Muhittin Mungan for reading the draft.

  \bibliographystyle{eplbib.bst}
  \bibliography{ken_LNP_sar.bib}

\revision{(Q1)\!\!}{
\section{Appendix A: Sign of $\psi_i$ and the direction of `time'\label{sec:app-sign}}
As noted in the footnote \ref{fn:A}, the assignment of the sign of $\psi_i$ is  linked to the assignment of the `time' direction, $i\to (i+1).$ 
When we rotate the whole system by the angle $\pi$ {\it within} the 2D plane,
we could assign the `time' index $i$ in the reverse order. 
In more detail, the same conformation of the system can undergo the following changes over $0\le i\le m-1$: 
$$
\inRbracket{ \begin{array}{lll} i \\    \psi_i  \\  \psi_{i+1}   \end{array}}
\mapsto 
\inRbracket{ \begin{array}{lll} m-1-i \\   - \psi_{m-1-i}  \\  -\psi_{m-1}  \end{array}}.
$$
This mapping involves three inversions, including the `temporal' one.
Then the global concave shape is mapped into global convex one, and vice versa.
Furthermore, a \red{stable} fixed point becomes \red{unstable  upon} this mapping.
While the domain of $\{\psi_i,\psi_{i+1}\}$ is moved from the first quadrant to the third one, the convexity of the return map, $\psi_i\mapsto \psi_{i+1}=\Phi(\psi_i)$ (see later), remains in the same sign.}\\

\revision{(Q2)\!\!}{
\section{Appendix B: Global shape variable and its change}\label{sec:app-shape}
While there is not a unique way to define the ``global shape variable" of the chain system,
we can use, for instance, the first flexion angle $\psi_0$ as the representative single degree of freedom.

 In Fig.\ref{fig:convergence-FP}(b), for example, this angle has been chosen at $\pi/12$ for the middle profile while the upper [lower] profiles have started with $\psi_0$ slightly above [below] this fixed-point angle, respectively. As we will see in Fig.\ref{fig:return-maps}(d-f) the return map tells that (i) the sequence $\{\psi_i\}$ is strictly monotonous, either increasing or decreasing, except for the fixed-point sequence, and (ii) a sequence $\{\psi_i\}$ never extends across the fixed-point values.  In other words, when the representative angle (eg. $\psi_0$) is continuously changed so that it meets and crosses a fixed-point value, all the members of $\{\psi_i\}$ do the same {\it simultaneously}. 
Except at the critical case (Figs.\ref{fig:return-maps}(b) and (e)), where the stable and unstable fixed points are merged, an increasing sequence before crossing a fixed point should become a decreasing sequence after the crossing, and vice-versa.
\footnote{\revision{(Q2)\!\!}{\, Please avoid confusion with the {\it module parameters}, which characterise the shape of the individual backbone module. The module parameters are fixed when we discuss the change of the global shape variable. By contrast, when we discuss the bifurcation, it is the module parameters that are varied. The latter variations cause the change in the topology of the return map through the bifurcation parameter ($\mu,$ see below).}}
}\\

\revision{(Q3+R2)\!\!}{
\section{Appendix C: Normal form from the return map (\ref{eq:map-cond})}\label{sec:app-NF}
In the main text, the return map from $\psi_i$ to $\psi_{i+1}$
is given through (\ref{eq:map-cond}). 
Below we will show how the normal form (\ref{eq:normal}) is explicitly derived from (\ref{eq:map-cond}) as a limiting case.
 For the simplicity of calculation, we rather deal with the function $K(x,y)$ such that the condition, $K(\psi_i,\psi_{i+1})=0,$ is equivalent to (\ref{eq:map-cond}): 
\small{\beqa 
\label{eq:Kxy}
&&
\! \! \! \! \! \! \! \! \! \! \! \! 
 K(x,y)
\equiv 
\!\left\|  \left(\!\!\! \begin{array}{c} 1\\ 0 \end{array}\!\!\!
 \right) \!+\!  f \! \left(\!\!\! \begin{array}{c} \cos(-\phi _0 \!+\!  {y})\\ \sin(-\phi _0 \!+\!  {y})  \end{array}\!\!\! \right) \!-\!  n \! \left(\!\!\! \begin{array}{c} \cos(\nu _0 \!-\!  {x} )\\ \sin(\nu _0 \!-\!  {x} )   \end{array}\!\!\! \right)  \right\|^2 
\! \! \! 
-\!  (L_{0})^2.
\cr &&
\eeqa}
By (\ref{eq:L0}) the length squared of the shaft reads 
$(L_{0})^2 =$ $1+f^2+n^2+2(f \cos(\phi_0)-n \cos(\nu_0)+fn \cos(\phi_0+\nu_0)).$
By construction $K(0,0)\!\!=\!\!0$ is assured, corresponding to the straight conformation.
In order to observe the essential behavior of the bifurcation we regard both $x$ and $y$ are small and develop $K(x,y)$ around $x=y=0$ up to the second order:
\beqa \label{eq:expK}
K(x,y) \!\!\!\!
&=& \!\!\!\!
K_{1,0}x+K_{0,1}y
+\inv{2}K_{2,0}x^2+K_{1,1}xy+\inv{2}K_{0,2}y^2
\cr && +\mathcal{O}(x^3,x^2 y,xy^2,y^3),
\eeqa
where 
$$
K_{n,m}\equiv \left.\frac{\partial^{n+m} K(x,y)}{\partial x^n \partial y^m}
\right|_{x=y=0}
$$
and their concrete forms are
\beq
\inRbracket{\begin{array}{c} 
K_{0,0}\\ K_{1,0}\\ K_{0,1}\\ K_{2,0}\\ K_{1,1}\\  K_{0,2}\\ \end{array}}
=\inRbracket{\begin{array}{c} 
0\\ -2fn [  \sin ({\nu_0}+{\phi_0})+ \sin   ({\nu_0})/f]\\
-2 fn [ \sin ({\nu_0}+{\phi_0})-\sin   ({\phi_0})/n]\\
 2fn [  \cos ({\nu_0}+{\phi_0})+ \cos({\nu_0})/n]\\
 2 f n \cos ({\nu_0}+{\phi_0})\\
 2fn [  \cos ({\nu_0}+{\phi_0})+ \cos({\nu_0})/f]\\ \end{array}}
\eeq
First we assess the qualitative feature near the trivial fixed point $\psi_{i+1}=\psi_i=0$ of the return map by considering  (\ref{eq:expK}) up to the first order terms. 
For this purpose we formally write $x=\epsilon x_1$ and $y=\epsilon y_1$ in (\ref{eq:expK}), and then solve $K(x,y)=0$ up to the first order of $\epsilon.$ 
We have 
$$
\psi_{i+1}=\inRbracket{-\frac{K_{1,0}}{K_{0,1}}}\psi_i+\mathcal{O}(\epsilon^2)
$$
The Figs.\ref{fig:return-maps}(d-f)  show that $-(K_{1,0}/K_{0,1})$ is the inclination, $d\psi_{i+1}/d\psi_i,$  at the origin. The case of $-(K_{1,0}/K_{0,1})=1$  is, therefore, the critical point, see Fig.\ref{fig:return-maps}(e). Likewise, the cases, $-(K_{1,0}/K_{0,1})<1$ and $-(K_{1,0}/K_{0,1})>1$ correspond, respectively, to Figs.\ref{fig:return-maps}(d) and (f).
In terms of the module parameters the critical shape condition $-(K_{1,0}/K_{0,1})=1$ must give the form of $\Lambda(f,\phi_0,n,\nu_0)=0$ in the main text \S{\bf Normal form}. Its explicit form then reads:
$$
\Lambda(f,\phi_0,n,\nu_0)=f \sin(\phi_0)-n \sin(\nu_0)-2 fn \sin(\phi_0+\nu_0).
$$
As discussed below (\ref{eq:normal}) the last condition defines the three-dimensional critical manifold in the four-dimensional shape \revision{}{parameter}  space $(f,\phi_0,n,\nu_0).$

Next we look for the solution of $K(x,y)=0$ up to the second order of $\epsilon$ but 
{\it at critical} condition $-(K_{1,0}/K_{0,1})=1.$ 
Substituting $x=\epsilon x_1$ and $y=\epsilon y_1+\epsilon^2 y_2$ into (\ref{eq:expK}),
the condition $K(x,y)=0$ reads, after coming back to the notations, $(\psi_i,\psi_{i+1})$ for $(x,y),$ as follows:
\beq \label{eq:criticalNF}
\psi_{i+1}=\psi_i+\lambda{{\psi_i}^2}+\mathcal{O}(\epsilon^3) \quad \mbox{(critical)},
\eeq
where 
\beqa \label{eq:lambda}
\lambda &\equiv& \frac{{K_{2,0}}{}+2K_{1,1}+\red{{K_{0,2}}{}}}{2K_{1,0}}
\cr 
&=&-\frac{{\cos(\nu_0)+2 f \cos(\phi_0+\nu_0)} }{{\sin(\nu_0)+f \sin(\phi_0+\nu_0)}}.
\eeqa
Because (\ref{eq:lambda}) has odd parity upon the simultaneous change $(\phi_0,\nu_0) \to (-\phi_0,-\nu_0),$ the sign of $\lambda$ would be inversed when the backbone modules were mirror-inverted, i.e., upside-down.
In Fig.\ref{fig:return-maps}(e) we have $\lambda>0$ \footnote{\revision{(Q3+R2)\!\!}{\,For example, when the backbone takes the shape of Fig.\ref{fig:design}, i.e., $(f,\phi_0,n,\nu_0)=(1/2,\pi/2,1/2,\pi/2)$ we find $\lambda =1.$} }   and $\lambda$ remains non-vanishing across the bifurcation.

Subsequentely the topological characteristic of the bifurcation, i.e., the transcritical feature, is found by allowing $(-{K_{1,0}}/{K_{0,1}})$ to be weakly off-critical:
$$ 
-\frac{K_{1,0}}{K_{0,1}}=1-\lambda \mu \,; \qquad \mu=\mathcal{O}(\epsilon),
$$
where the condition $\mu=\mathcal{O}(\epsilon)$ ensures the proximity to the bifurcation. 
Then (\ref{eq:criticalNF}) is extended to the weak off-critical case as follows:
\beq \label{eq:off-criticalNF}
\psi_{i+1}=\psi_i-\lambda\psi_i (\mu -{\psi_i})+\mathcal{O}(\epsilon^3).
\eeq
As for $\lambda,$ which is of order $\epsilon^0=1,$ we can use $\lambda$ of (\ref{eq:lambda}) in the weak off-critical regime within the error of $\mathcal{O}(\epsilon^3)$.\footnote{\revision{}{We can verify this by inserting formally $\lambda=\lambda_0+\epsilon \lambda_1$ into (\ref{eq:off-criticalNF}).}}
Altogether we reach the normal form of transcritical bifurcation (\ref{eq:normal}) as the structure of the mapping (\ref{eq:map-cond}), or $K(\psi_i,\psi_{i+1})=0$ with (\ref{eq:Kxy}), near the bifurcation point ($\mu=\mathcal{O}(\epsilon)$) and near the straight conformation ($\psi_i=\mathcal{O}(\epsilon)$).}

\end{document}